\documentstyle[12pt,epsfig]{article}        
\newlength{\dinwidth}                       
\newlength{\dinmargin}                      
\def\lsim{\mathrel{\rlap{\lower4pt\hbox{\hskip1pt$\sim$}}
    \raise1pt\hbox{$<$}}}                
\def\gsim{\mathrel{\rlap{\lower4pt\hbox{\hskip1pt$\sim$}}
    \raise1pt\hbox{$>$}}}                
\newcommand{\bq}{\begin{equation}}
\newcommand{\eq}{\end{equation}}

\newcommand\GeV{\,\mbox{GeV}}

\newcommand\MP{M^2_{\Phi}}

\newcommand\sh{\hat{s}}

\begin{document}
\begin{titlepage}

\large
\normalsize
\begin{flushleft}
August 1996
\end{flushleft}
\vspace*{4cm}
\begin{center}
\LARGE
{\bf Leptoquark Pair Production at HERA} \\
\vspace{3cm}
\large
Johannes Bl\"umlein$^a$, Edward Boos$^{a,b}$, 
and Alexander Kryukov$^{a,b}$
\\
\vspace{2.5cm}
\large {\it
$^a$ DESY--Zeuthen \\

\vspace{0.1cm}
Platanenallee 6, D--15735 Zeuthen, Germany }\\
\vspace{1.0cm}
\large {\it
$^b$ Institute of Nuclear Physics, Moscow State University,
RU--119899 Moscow, Russia} \\
\vspace{\fill}
\normalsize
{\bf Abstract} \\
\end{center}
\noindent
The scalar and vector leptoquark pair production cross sections for  deep
inelastic $ep$ scattering are calculated. Estimates are presented for the
search potential at HERA.
\vfill 
\noindent
\normalsize
\begin{center}
{
\sf
Contribution to the Proceedings of the 1996 HERA Physics
Workshop}
\end{center}
 
\end{titlepage}
\vspace*{1cm}
\begin{center}  \begin{Large} \begin{bf}
 Leptoquark Pair Production at HERA \\
  \end{bf}  \end{Large}
  \vspace*{5mm}
  \begin{large}

Johannes Bl\"umlein$^a$, Edward Boos$^{a,b}$, and
Alexander Kryukov$^{a,b}$
  \end{large}

$^a$ DESY--Zeuthen,
     Platanenallee~6,~D-15735~Zeuthen, Germany\\
$^b$ Institute of Nuclear Physics, Moscow State University,
RU--119899 Moscow, Russia \\
\end{center}
\begin{quotation}
\noindent
{\bf Abstract:}
The scalar and vector leptoquark pair production cross sections
 for  deep
inelastic $ep$ scattering are calculated. Estimates are presented for the
search potential at HERA.
\end{quotation}

\section{Introduction}
\label{sect1}
\noindent
In many extensions of the Standard Model bosonic states carrying
both lepton and quark quantum numbers, so--called
leptoquarks,  are contained.
Leptoquarks may exist in the mass range reached by high energy colliders
if their couplings are $B$ and $L$ conserving. A general classification
of these states was given in ref.~\cite{BRW} demanding also
non--derivative and family diagonal couplings.
In most of the scenarios
the fermionic leptoquark couplings are not predicted.
Moreover,
a detailed analysis of low energy data~\cite{LEUR}
showed that the these leptoquark  couplings  are small in the mass range
up to $O(1~{\rm TeV})$.
Thus processes depending on the fermionic
couplings can not be used to obtain rigorous mass bounds for these
states.

On the other hand, the  couplings of the leptoquarks  to the electroweak
gauge bosons and gluons are determined by the respective gauge
symmetries. In the case of scalar leptoquarks the couplings are thus
completely predicted.
For vector leptoquarks additionally anomalous couplings may contribute.
Due to the small fermionic couplings the pair production cross sections
depend only on the bosonic couplings and mass limits may be derived
directly.

In the present paper a brief account is given on results obtained
in refs.~\cite{JB1,JB2} and  estimates are presented for the search
potential in the HERA energy range.

\section{The Pair Production Cross Sections}
\label{sect2}
\noindent
The integral
leptoquark pair production cross sections in deep inelastic $ep$
collisions are described by
\begin{equation}
\sigma_{S,V}^{ep,tot} = \sigma_{S,V}^{ep,dir} + \sigma_{S,V}^{ep,res},
\end{equation}
containing a direct and a resolved photon contribution which are
given by
\begin{equation}
\sigma_{S,V}^{ep,dir} = \int_{y_{min}}^{y_{max}} dy
                        \int_{x_{min}}^{x_{max}} dx \phi_{\gamma/e}(y)
                        G_{p}(x,\mu^2)
\hat{\sigma}_{S,V}^{dir}(\hat{s},M_{\Phi})
\theta(\hat{s} - 4M^2_{\Phi}),
\end{equation}
and
\begin{eqnarray}
\label{xsep}
\sigma_{S,V}^{ep, res}(s,M_{\Phi}) &=&
\int_{y_{min}}^{y_{max}} dy
\int_{4 \MP/Sy}^1 dz
\int_{4 \MP/Syz}^1 dx
\phi_{\gamma/e}(y)
\theta(\hat{s} - 4 M_{\Phi}^2)
\nonumber\\
&\times&
\left \{
\sum_{f=1}^{N_f} \left [ q_f^{\gamma}(z, \mu_1)
\overline{q}_f^p(x,\mu_2) +
 \overline{q}_f^{\gamma}(z, \mu_1)
q_f^p(x,\mu_2) \right ]
                 \hat{\sigma}_{S,V}^q(\sh, M_{\Phi})
\right. \nonumber\\
&+& \left.
G^{\gamma}(z, \mu_1) G^p(x,\mu_2)
                 \hat{\sigma}_{S,V}^g(\sh, M_{\Phi})
 \right \},
\end{eqnarray}
respectively.
Here $\phi_{\gamma/e}$ denotes the Weizs\"acker--Williams distribution
and $M_{\Phi}$ is the leptoquark mass.
$q_f^{\gamma}$ and $G^{p(\gamma)}$ are the quark and gluon distributions
in the photon and proton, respectively, $\hat{s} = S x y$,
and $\mu_1$ and $\mu_2$
denote the factorization scales.

The subsystem
scattering cross sections $\hat{\sigma}_{S,V}^{q,g}(\hat{s}, M_{\Phi})$
were calculated in \cite{JB1} for the direct process and in \cite{JB2}
for the resolved processes, both for scalar and vector leptoquarks.
There also the differential scattering cross
were derived. In the case of vector leptoquarks the scattering cross
sections were calculated accounting both
for anomalous photon $\kappa_A, \lambda_A$, and gluon  couplings,
$\kappa_G, \lambda_G$. These contributions are understood
in an effective description being valid
in the threshold range, i.e. for
$S \sim 4 M_{\Phi}^2$. Due to the anomalous couplings the pair production
cross sections for vector leptoquarks obtain as well unitarity violating
pieces which however are
assumed to never become large. 
It
is hardly possible in  general,
to provide a
correct high energy description in a
model--independent way,
as intended in the present paper focussing on the threshold range only.
This, instead,
requests to consider  a specific
scenario accounting
also for the details of
the respective pattern of symmetry
breaking.

For all
details of the calculation we refer to
refs.~\cite{JB1} and \cite{JB2}.
\section{Numerical Results}
\label{sect3}
\noindent
In figures~1 and 2 the integrated scattering cross sections for a
series of scalar and vector leptoquarks are shown in dependence of
the leptoquark mass and charges. For the vector leptoquarks different
choices of anomalous couplings are also considered. For simplicity we
identified $\kappa_A = \kappa_G$ and $\lambda_A = \lambda_G$.
It is interesting to note that not the Yang--Mills type couplings,
$\kappa = \lambda = 0$, but the so--called  minimal couplings,
$\kappa = 1,
\lambda = 0$, result in the smallest cross section. In further
experimental studies it might be interesting to vary even all the four
anomalous
couplings independently. As seen in figures~1 and 2 the integral cross
sections behave about like
\begin{equation}
\label{SIG}
\sigma_{tot}^{S,V}(M_{\Phi}) \sim A \exp(-B M_{\phi}).
\end{equation}
This relation can be used to obtain an estimate of the respective search
limits which can be reached at a given integrated
luminosity, ${\cal L}$.

For ${\cal L} = 100~{\rm pb}^{-1}$  and $\sqrt{s} = 314 \GeV$
the search limits
for charge $|Q_{\Phi}| = 5/3$
scalar leptoquarks ranges up to
$60~(45)~{\rm GeV}$ and for
vector leptoquarks up to $70~(55)~{\rm GeV}$,
given a
signal sample of 10 (100) events,
respectively.

For most of the channels the experiments at LEP~1 have
excluded leptoquarks with masses below $M_Z/2$.  At present the
most stringent mass bound for both scalar~\cite{TEVA}
 and vector leptoquarks\footnote{Studies considering also
anomalous leptoquark couplings were not performed yet.}
decaying into the fermions of the first
and second
 family come from
TEVATRON and exclude the range $M_{\Phi} \lsim 90~{\rm GeV}$.
For some leptoquark types the
range $M_{\Phi} \lsim 130~{\rm GeV}$ is excluded~\cite{TEVA}.
No bounds were yet derived  for
3rd generation leptoquarks, e.g. those decaying  as
$\Phi_{S,V} \rightarrow b~\tau$,~etc.,
in the TEVATRON analyses. Due to the
lower background rates, an investigation of particularly
this channel may be more
suited to $ep$ or $e^+e^-$ collisions than for proton collisions.


\newpage
\begin{center}

\mbox{\epsfig{file=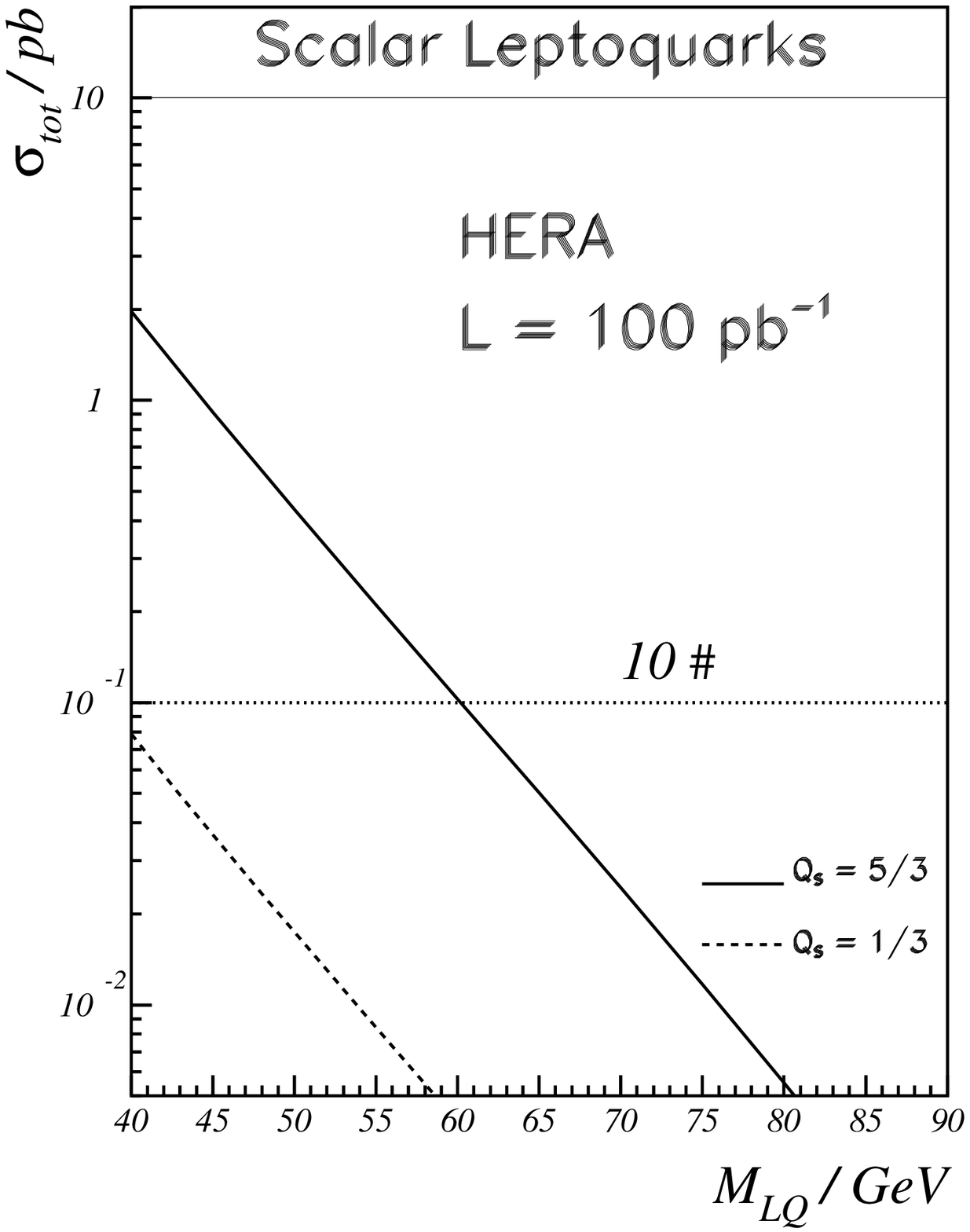,height=18cm,width=16cm}}

\vspace{2mm}
\noindent
\small
\end{center}
{\sf
Figure~1:~Integrated cross
sections for scalar leptoquark pair production at HERA,
$\sqrt{S}~=~314~\GeV$.
Full line:~$\sigma_{tot}$ for $|Q_{\Phi}| = 5/3$;
dotted line:~$\sigma_{dir}$ for $|Q_{\Phi}| = 5/3$;
dashed line:~$\sigma_{tot}$ for $|Q_{\Phi}| = 1/3$;
dash--dotted line:~$\sigma_{dir}$ for $|Q_{\Phi}| = 1/3$.
}
\normalsize
\newpage
\begin{center}

\mbox{\epsfig{file=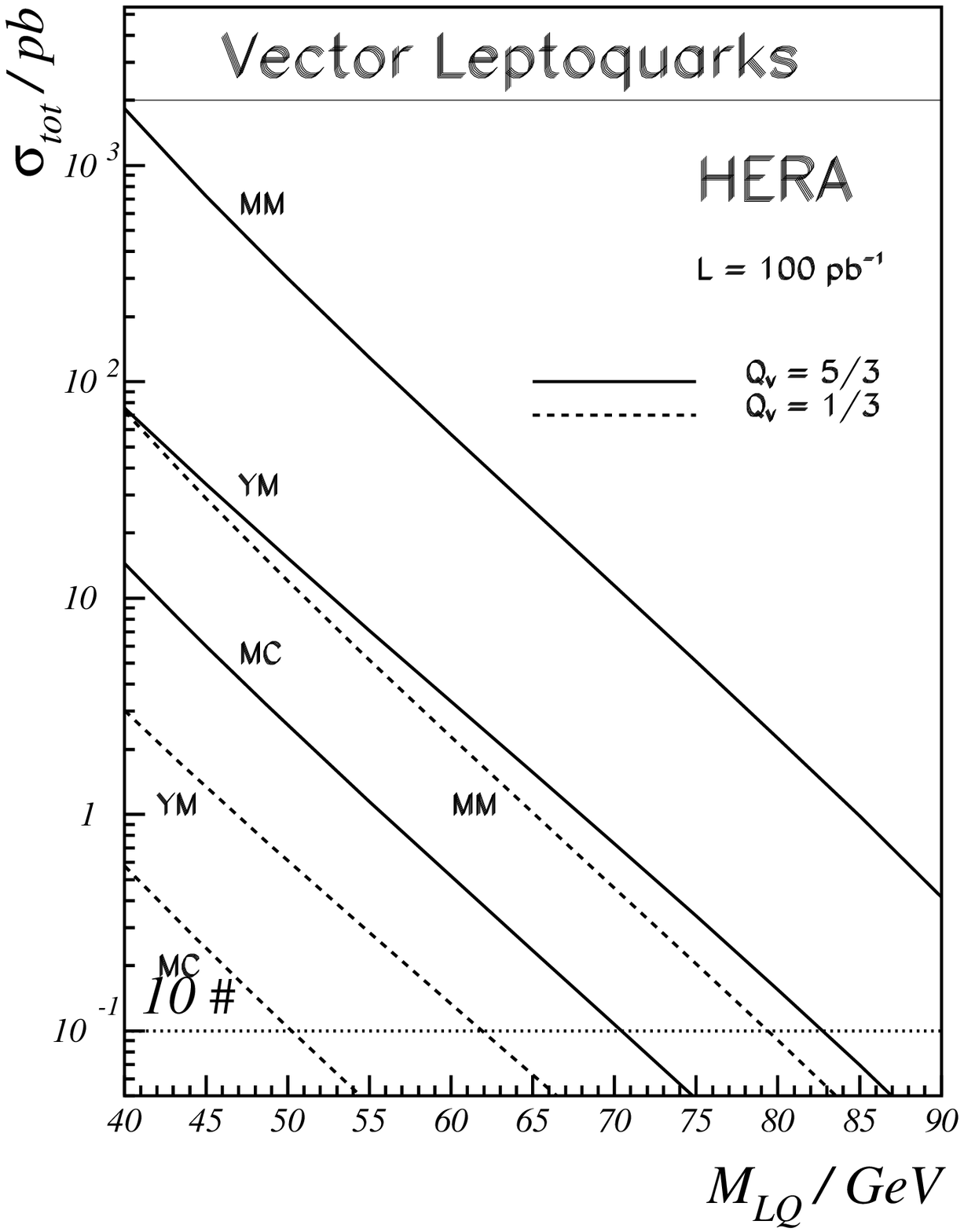,height=18cm,width=16cm}}

\vspace{2mm}
\noindent
\small
\end{center}
{\sf
Figure~2:~Integrated cross
sections $\sigma_{tot} = \sigma_{dir} + \sigma_{res}$
for vector leptoquark pair production at HERA,
$\sqrt{S}~=~314~\GeV$.
Upper full line:~$|Q_{\Phi}| = 5/3, \kappa_{A,G} = \lambda_{A,G} = -1$
(MM5);
Upper dashed line:~$|Q_{\Phi}| = 5/3, \kappa_{A,G} = \lambda_{A,G} = 0$
(YM5);
Upper dotted line:~$|Q_{\Phi}| = 5/3, \kappa_{A,G} = 1,
\lambda_{A,G} = 0$  (MC5). The corresponding lower lines are those
for $|Q_{\Phi} = 1/3$.
}
\normalsize
\end{document}